**Sacrificial Template Replication: Fabrication of arbitrary three-dimensional suspended hollow microstructures in transparent fused silica glass**

*Frederik Kotz[1], Patrick Risch[1], Karl Arnold[1], Alexander Quick[2], Michael Thiel[2], Andrei Hrynevich[3], Paul D. Dalton[3], Dorothea Helmer[1], Bastian E. Rapp[1]\**

1       NeptunLab, Institute of Microstructure Technology (IMT), Karlsruhe Institute of Technology (KIT), Hermann-von-Helmholtz-Platz 1, Eggenstein-Leopoldshafen, 76344, Germany.

2       Nanoscribe GmbH, Hermann-von-Helmholtz-Platz 1, Eggenstein-Leopoldshafen, 76344, Germany.

3       Department for Functional Materials in Medicine and Dentistry and Bavarian Polymer Institute, University Würzburg, Pleicherwall 2, Würzburg, 97070, Germany.

\*       Correspondence to: Bastian E. Rapp, Bastian.Rapp@kit.edu

## Abstract

Fused silica glass is the preferred material in all applications which require long-term chemical and mechanical stability as well as excellent optical properties. The manufacturing of complex hollow microstructures within transparent fused silica glass is of high interest for, among others, the miniaturization of chemical synthesis towards more versatile, configurable and environmentally friendly flow-through chemistry as well as high-quality optical waveguides or capillaries. However, microstructuring of such complex three-dimensional structures has proven evasive due to the high thermal and chemical stability as well as mechanical hardness of glass. Here we present an approach for the generation of hollow microstructures in fused silica glass with unseen precision and freedom of three-dimensional designs. The process combines the concept of sacrificial template replication with a room-temperature molding process for fused silica glass. The fabricated glass chips are versatile tools for, among other, the advance of miniaturization in chemical synthesis on chip.

## Acknowledgments

This work has been partly funded by the German Federal Ministry of Education and Research (BMBF), "Fluoropor", funding code: 03X5527. We thank Evonik for providing nanopowders. We thank Richard Thelen for white light interferometry. This work was partly carried out with the support of the Karlsruhe Nano Micro Facility (KNMF) (www.kit.edu/knmf), a Helmholtz Research Infrastructure at Karlsruhe Institute of Technology (KIT) (www.kit.edu).

# 1 Introduction

Microstructures in fused silica glass are usually fabricated by wet chemical or dry etching processes [1]. More complex structures can be fabricated using precision glass molding, sol-gel replication or powder blasting [2-4]. However all these techniques are only capable of fabricating open, two-dimensional channel structures which need to be bonded with a planar substrate to fabricate simple suspended hollow microstructures such as microfluidic channels. Creating freeform hollow structures inside fused silica glass is difficult and has, until now, only been shown using femtosecond laser writing with consecutive etching of the irradiated areas with aggressive chemicals such as hydrofluoric acid (HF) [5, 6]. However for long channel structures with few inlets the etching is inhomogeneous and results in tapered channel structures with significantly wider dimensions towards the channel inlets [7]. Additionally, channel lengths are limited by the etching process, since HF etching shows a decrease in etching speed over channel length and debris can quickly block the channels [8]. To overcome the problem of varying channel diameters, different techniques such as wobbling or drawing have been established [9, 10]. However, these techniques can only be employed for simple channel geometries. To overcome the need for aggressive etching solutions, femtosecond laser writing by liquid-assisted ablation was developed [11, 12]. This technique generates components with significant surface roughness which require post-treatment to generate surfaces of optical quality [5, 13]. A comparison of relevant techniques to structure fused silica glass can be found in Table 1.

Table 1: Comparison of relevant methods for structuring fused silica glass.

| Method | 3D capability | Suitable for 3D microvoids (1 – 100 µm) | Resolution | Surface quality | Literature |
|---|---|---|---|---|---|
| **Etching** | | | | | |
| Wet chemical etching | -- | -- | ~ 1 µm | ~ 1-10 nm (Ra) | [14, 15] |
| Dry etching | -- | -- | < 1 µm | 0.5(rms)-2 nm (Ra) | [16, 17] |
| **Mechanical** | | | | | |
| Powder blasting | -- | -- | > 10 µm | 0.1 – 10 µm (Ra) | [18] |
| **Laser-assisted** | | | | | |
| Laser-assisted etching | + | + | 1-2 µm | 0.1-0.2 µm (rms) | [19-21] |
| Backside etching | -- | -- | 2 µm | 0.05– 0.5 µm | [22] |
| **Replication** | | | | | |
| Sol-Gel | -- | - | < 1 µm | n.a. | [3] |
| Nanocomposites | -- | - | < 1 µm | 2 nm (rms) | [23, 24] |
| Precision glass molding | -- | -- | ~ 1 µm | 2 nm | [2, 25] |
| **Additive** | | | | | |
| Stereolithography nanocomposites | ++ | - | 60 µm | 2 nm (rms) | [26] |
| Sol-Gel | ++ | -- | 200 µm | n.a. | [27] |
| Stop flow lithography | -- | -- | 10 µm | 6 nm (rms) | [28] |

As of today there is no method for generating truly arbitrary three-dimensional hollow structures of centimeter lengths and few micrometers diameter in bulk fused silica glass. However, many applications such as, e.g., microfluidics, flow-through synthesis, photonics or waveguiding applications in optics and photonics require methods for creating freeform hollow structures in fused silica with smooth surfaces. These structures are also highly sought for flow-through on-chip chemical synthesis, a field which has recently gained significant attention [29, 30]. Miniaturization of chemical reactions promises significantly reduced reactant consumption, more stable reaction conditions and new reaction pathways such as ultrafast mixing or kinetic reaction control which are inaccessible in standard batch and flask chemistry [31, 32]. Recently, the use of additive manufacturing for the manufacturing of configurable, low-volume synthesis systems has gained significant attention [33]. However, suitably high resolution structures can, until now, only be manufactured in polymers, which limits the choice of solvents, temperature and pressure. In many applications the established chemistries (optimized for glassware) cannot be directly translated to polymers and significantly lower reaction yields and efficiencies have been achieved. In a recent contribution to *Science*, Kitson et al., stated that for polymer-based reactionware suitable for on-demand pharmaceutical synthesis, a translation process from glassware to polymers would be necessary [34]. As glass is the material of choice to withstand the harsh reaction conditions as well as to enable the on-line analysis of reactions through spectroscopy, novel approaches to manufacturing of intricate three-dimensional glass structures are highly sought.

## 2 Sacrificial Template Replication

We have recently developed a method for structuring fused silica components at room temperature [23, 24, 26]. In this process, a nanocomposite consisting of a high amount of fused silica nanoparticles in an organic binder matrix is polymerized at room temperature and consecutively sintered to full-density, transparent fused silica glass. We have shown that using stereolithography three-dimensional fused silica glass structures can be fabricated. Fabrication of suspended three-dimensional hollow microstructures by 3D printing however remains intrinsically difficult since entrapped uncured material inside the microvoids is difficult to remove and is partially cured during the printing process, thereby blocking the microstructures. Here we demonstrate that a combination of the casting of these nanocomposites and sacrificial template replication (STR), a concept known from polymer and ceramic processing, is able to produce complex, suspended hollow microstructures in fused silica glass. In sacrificial template replication, a template structure or immiscible phase is introduced into a material and consecutively removed by dissolving, etching or burning to produce a desired hollow structure [35-39]. Nanochannels in glass can be produced by coating of electrospun nanofibers with silicon dioxide and consecutive calcination, but have so far not been shown for fused silica and are restricted to single electrospun fibers and therefore limited to very simple designs [40]. Sacrificial template techniques have been used in combination with metallic glasses and bioglasses. However, the structures produced are restricted to open structures with lamellae or fibrous/porous surface structures [41, 42]. By combining room temperature glass structuring and sacrificial template replication, it is possible, for the very first time, to generate nearly arbitrarily-shaped freeform three-dimensional channels and hollow structures in fused silica glass.

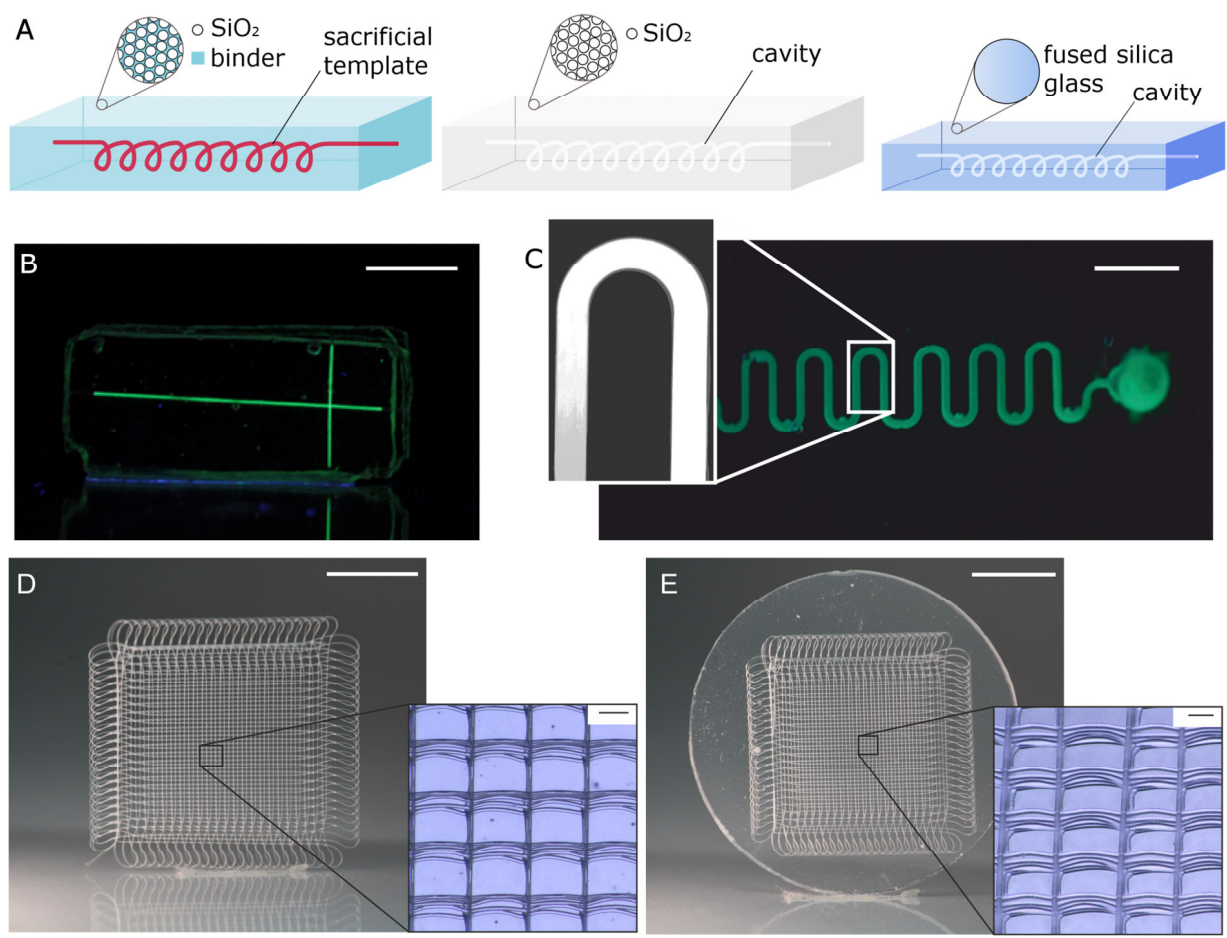

*Figure 1*: Fabrication of suspended hollow microstructures in fused silica glass using sacrificial template replication (STR): A) Polymeric filaments are embedded in an amorphous silica nanocomposite. The polymerized nanocomposite is turned into fused silica glass via thermal debinding and sintering. The polymeric template is removed and leaves the according hollow cavity. B) Microfluidic fused silica chip fabricated by embedding a nylon thread (scale bar: 9 mm). C) Microfluidic meander fabricated by embedding polymerized PEGDA structured by microlithography (scale bar: 11 mm). D) A mesh structure made from poly(ε-caprolactone) using melt electrowriting (scale bar: 5 mm). The inset shows the microscopy image of the mesh with a fiber diameter of 25.0 µm (scale bar: 100 µm). E) Inverse hollow mesh structure in fused silica glass (scale bar: 4.5 mm). Inset shows the microcavities with a width of around 18.4 µm (scale bar: 100 µm).

## 3    Results and Discussion

We show template molding using nylon threads, poly(ethylene glycol diacrylate) (PEGDA) scaffolds, poly(ε-caprolactone) (PCL) microfiber meshes produced by melt electrowriting and complex polymeric microstructures fabricated by direct laser writing. All scaffolds were immersed in the nanocomposite and consecutively processed to decompose and evaporate polymeric residue and to give fused silica glass microstructures. As the templates are removed in the gas phase, there is no material redeposition or channel blocking by incomplete removal. Also diffusion limitations which usually restrict the length and dimensions of etched structures, do not apply. The STR process in fused silica glass is shown in Figure 1A. First the polymeric template is embedded in the nanocomposite. The nanocomposite is then polymerized using light exposure. Thermal debinding of the polymeric binder and the template is done at 600 °C and ambient pressure. The structures are consecutively sintered at a pressure of 5 x $10^{-2}$ mbar at 1300 °C (see Table 2 for the optimized protocol for thermal debinding and sintering).

A significant benefit of this nanocomposite approach is that it is compatible with different materials and fabrication processes for the sacrificial templating component. Simple microfluidic channels were fabricated by immersing nylon threads in the nanocomposite (see Figure 1B). Multiple threads were connected by thermally fusing the threads under light pressure at 100 °C. More complex two-dimensional microfluidic channels were fabricated by microlithography using PEGDA as material for the template (see Figure 1C).

Another accessible additive manufacturing approach for the templates is melt electrowriting which creates continuous, ultrafine diameter fibers [43, 44]. The smooth, uniform fibers produced in this process are well-suited as templates for generating structures such as complex microfluidic channel networks. Figure 1D shows a three-dimensional mesh structure of well-stacked melt electrowritten fibers. The mesh was completely immersed in the nanocomposite and processed/sintered as a bulk structure to give suspended channels in one step. The inverse micromesh structure obtained in fused silica glass is shown in Figure 1E. The glass cavities have a diameter of 18.4 μm.

In order to generate highly complex template microstructures direct laser writing was used. Figure 2 shows exemplary microstructures fabricated using this approach. The feature resolution achievable with this process is unmatched and comparably complex microstructures have never been manufactured in fused silica glass.

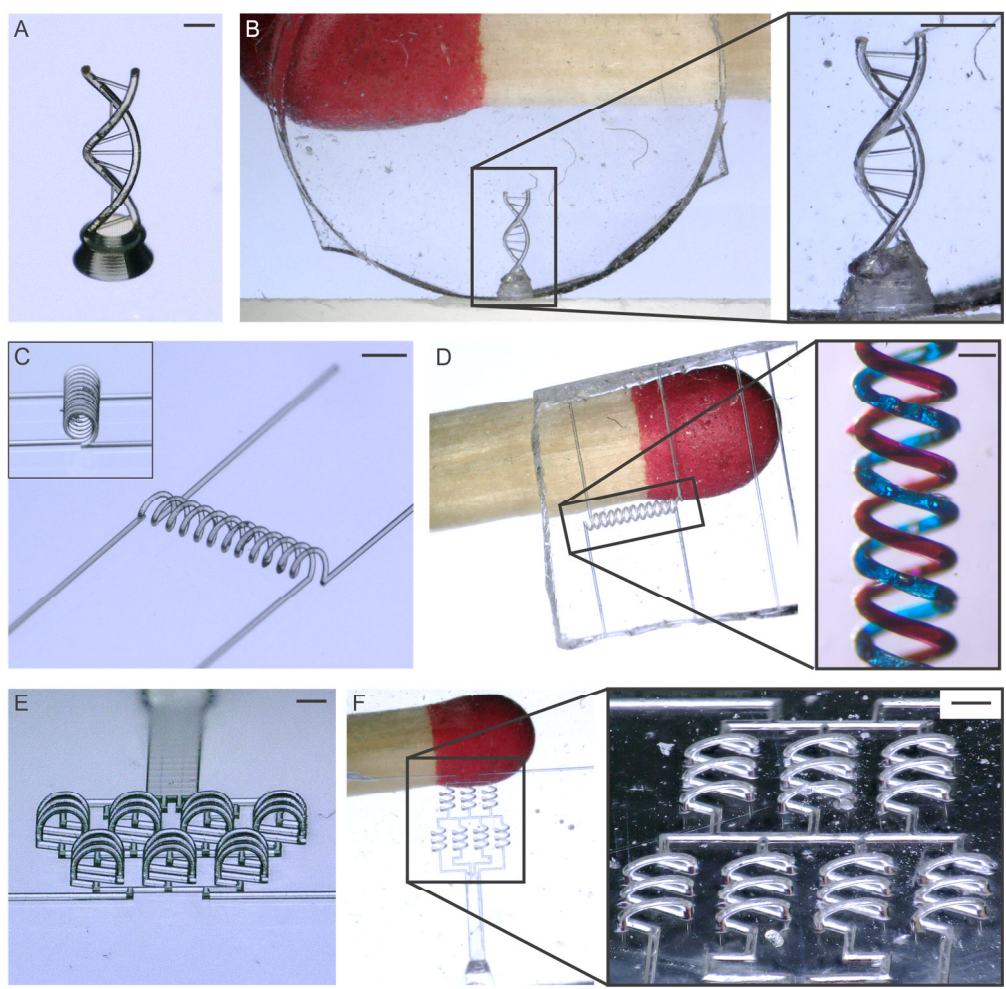

**Figure 2**: Suspended hollow microstructures in fused silica fabricated using sacrificial template replication and direct laser writing for the template generation. A) Polymeric DNA double helix (scale: 500 μm). B) Inverse structure in fused silica glass (scale: 400 μm). The smallest channel size was 20 μm. C) Intertwined spirals (scale: 900 μm). D) Resulting intertwined microfluidic spiral channels in fused silica glass with a channel width of 74 μm. The channels were filled with dyes (see inset, scale: 140 μm). E) Polymeric microstructures of an out-of-plane mixer structure (scale: 600 μm). F) Microfluidic mixer structure in fused silica glass with a channel width of 74 μm (scale bar: 280 μm).

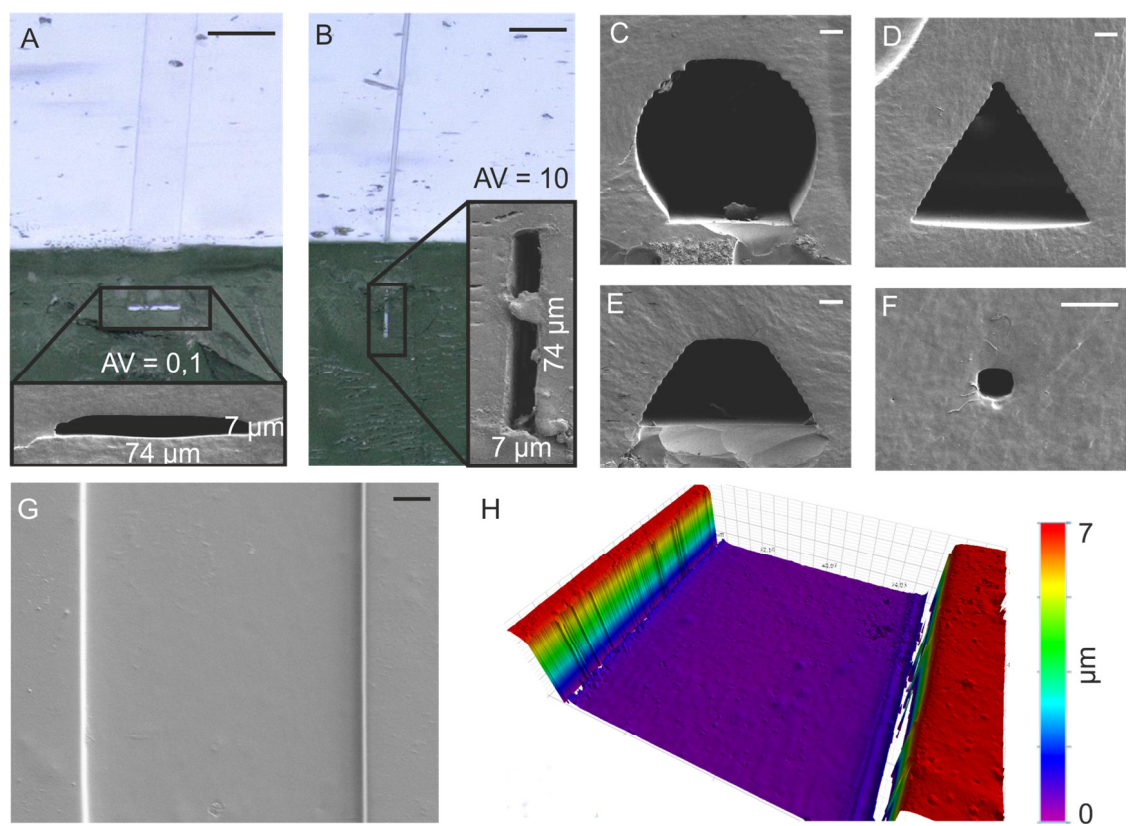

*Figure 3*: Characterization of suspended hollow microstructures in fused silica fabricated using sacrificial template replication and direct laser writing for template generation. A/B) SEM of rectangular channel cross section with an aspect ratio of 0.1 and 10 (scale: 100 µm). C-F) SEM of spherical, triangular, trapezoidal and rectangular channel cross-sections (scale: 10 µm). G/H) SEM and white light interferometry of the channel structure from A with a mean roughness of Ra ~ 20 nm (scale: 10 µm).

As an example, we generated a DNA double-helix structure (Figure 2A/B), intertwined microfluidic spiral channels (Figure 2C/D) and a three-dimensional microfluidic channel structure with out-of-plane geometries (Figure 2E/F).

Using this approach microfluidic channel structures with low and high aspect ratios can be fabricated with ease. Figure 3A/B shows two rectangular cross sections with an aspect ratio of 0.1 and 10, respectively. Furthermore, the cross sections of the channel structures can be changed to arbitrary shapes: Figure 3C-E shows circular, trapezoid and triangular channel cross sections with a maximum feature size of 74 µm. Channels of exceptionally small sizes can be fabricated as shown exemplary in Figure 3F, showing a microchannel of 7 × 7 µm² size. While the channel cross sections are in the range of only a few micrometers, the lengths of the channels are in the range of centimeters. The surface roughness of the produced fused silica depends on the type of mold used, i.e., the roughness of the polymeric template structure. The exceptionally smooth surfaces of the channels resulting from sacrificial templates fabricated by laser direct writing in fused silica glass is shown in Figure 3G/H showing mean roughness of $R_a$ ~ 20 nm. Surfaces of optical quality are therefore achievable.

## 4  Summary

In summary we described a novel potent technique to fabricate arbitrary embedded freeform three-dimensional suspended hollow microstructures in transparent fused silica glass by using a sacrificial template replication process. This technique will enable numerous applications in flow-through synthesis and analysis, microfluidics and Lab-on-a-Chip devices for chemical miniaturization as well as applications in optics and photonics.

## 5 Materials and Methods

*Materials:* Amorphous silica nanopowder of type Aerosil OX50 was kindly provided by Evonik, Germany. Hydroxyethylmethacrylate (HEMA) was purchased from Alfa Aesar, Germany. Tetraethylenglycoldiacrylate (TEGDA), polyethylenglycoldiacrylate 550 (PEGDA-550). phenylbis(2,4,6-trimethylbenzoyl)phosphine oxide, propylene glycol methyl ether acetate (PGMEA) and 2.2-dimethoxy-2-phenylacetophenone (DMPAP) were purchased from Sigma Aldrich. Negative-tone photoresists IP-S was purchased from Nanoscribe, Germany. 2-propanol was purchased from Carl Roth, Germany.

*Direct laser writing:* Prior to usage in the fabrication process, glass substrate (25 mm x 25 mm x 0.7 mm, from Nanoscribe GmbH, Germany) were activated by oxygen plasma in order to enhance the adhesion of the photoresist to the glass. Note that additional silanization is not recommended since the bonding was found to be too strong to detach the polymer from the substrate when embedded in the nanocomposite. The 3D objects were fabricated using a commercial lithography system Photonic Professional GT (Nanoscribe GmbH, Germany). Negative-tone photoresist IP-S was used as photoresist and was drop-casted on the activated substrate. The writing speed was set to 100 mm/s with the slicing distance set to 1 µm ("IP-S recipe" in software Describe, Nanoscribe GmbH, Germany). Both, solid writing and core-shell approach have been explored successfully. After exposure, the sample was developed in PGMEA for 10 min and rinsed with another bath of PGMEA for 30 seconds.

*Lithography:* PEGDA-550 was blended with 0.5 m% of the photoinitiator phenylbis(2,4,6-trimethylbenzoyl)phosphine oxide. PEGDA550 was then structured using a lithography system which we have previously described [45]. Structuring was done at a wavelength of 365 nm for 28 s at an exposure intensity of 2.6 mW cm$^{-2}$. After the exposure the polymeric structures were developed in 2-propanol for 30 s.

*Transfer of microstructures:* The microstructures were fabricated on a glass slide and the nanocomposite was cast on top. After polymerization, the nanocomposite with the embedded microstructure was peeled off the glass and the open structure was sealed with a second layer of nanocomposite. The process is shown in Figure 4.

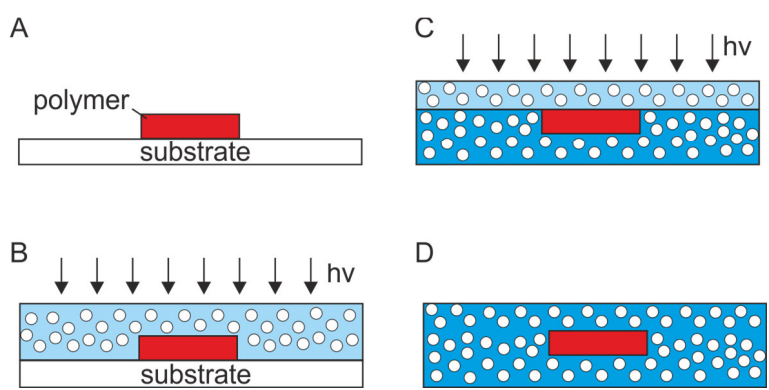

**Figure 4**: *Embedding of microstructures into the silica nanocomposite: A) Template structuring on a glass substrate. B) Casting and curing of the nanocomposite on the glass substrate. C) Closing the bottom of the structure by polymerizing a second layer of the liquid nanocomposite onto the polymerized nanocomposite. D) Embedded polymeric structure.*

*Melt electrowriting:* PCL (PC-12, Corbion, the Netherlands) was used as received and processed using a custom-built melt electrowriting printer as previously described [46]. Briefly, approximately 1 g of the PCL was placed in an electrically heated (75 °C) syringe and pneumatically delivered to a 23G nozzle using air (1.0 bar). This nozzle is positioned 6 mm above a collector, and a total of 5.5 kV is applied across this collector distance. Direct writing was performed using x/y linear stages and samples were used as sacrificial templates without post-processing.

*Preparation of the nanocomposite:* The nanocomposite used in this work consisted of 68 vol% HEMA, 7 vol% of TEGDA and 25 vol% of POE which were mixed prior to the dispersion process [23]. Afterwards 40 vol% Aerosil OX50 were dispersed in the monomeric mixture. The nanopowders were added in small increments to this mixture using a laboratory dissolver (R 130, IKA, Germany). Afterwards 0.5 m% (referred to the amount of reactive monomer) of the photoinitiator DMPAP was added following a further dispersion step of 30 min. Entrapped air bubbles were removed using a desiccator and a vacuum pump.

*Embedding and polymerization of the nanocomposite:* For embedding of the polymer filaments into the nanocomposites the latter were heated to 60 °C prior to the casting process. This reduces the risk of entrapping air bubbles. The nanocomposites were subsequently polymerized at a wavelength of 300-400 nm at an exposure intensity of 12 mW cm$^{-2}$ for 2 min.

*Heat treatment:* Thermal debinding was done using an ashing furnace (type AAF, Carbolite/Gero, Germany). Sintering was done using a tube furnace (type STF16/450, Carbolite/Gero, Germany) at a temperature of 1300 °C and a pressure of 5×10$^{-2}$ mbar with a heating rate of 3 K min$^{-1}$. The parameters for thermal debinding and sintering can be found in Table 2.

*Table 2: Thermal debinding and sintering protocol.*

| Program | Temperature [°C] | Heating rate [K/min] | Holding phase [min] |
|---|---|---|---|
| | 150 | 0,5 | 120 |
| Thermal debinding | 320 | 0,5 | 240 |
| | 600 | 0,5 | 120 |
| | 25 | 5 | - |
| | 800 | 3 | 90 |
| Sintering | 1300 | 3 | 120 |
| | 25 | 5 | - |